\documentclass[10pt,conference]{IEEEtran}
\IEEEoverridecommandlockouts
\usepackage{cite}
\usepackage{amsmath,amssymb,amsfonts}
\usepackage{algorithmic}
\usepackage{graphicx}
\usepackage{textcomp}
\usepackage{xcolor}
\usepackage{comment}
\def\BibTeX{{\rm B\kern-.05em{\sc i\kern-.025em b}\kern-.08em
    T\kern-.1667em\lower.7ex\hbox{E}\kern-.125emX}}
\begin{document}

\title{Empirical Analysis of Pull Requests for Google Summer of Code}

\author{\IEEEauthorblockN{Saheed Popoola}
\IEEEauthorblockA{\textit{School of Information Technology} \\
\textit{University of Cincinnati}\\
Cincinnati, USA \\
saheed.popoola@uc.edu}

}

\maketitle

\begin{abstract}
Internship and industry-affiliated capstone projects are popular ways to expose students to real world experiences and bridge the gap between academic training and industry requirements. However, these two approaches often require active industry collaboration, and many students struggle to find industry placements. Open-source contributions are a crucial alternative to gain real world experience, earn publicly verifiable contribution with real-world impact, and learn from experienced open-source contributors. The Google Summer of Code (GSoC) is a global initiative that matches students or new contributors with experienced mentors to work on open-source projects. The program aims to introduce the students to open-source development, help them gain valuable skills under the guidance of mentors, and hopefully encourage them to continue contributing to open-source projects. The realization of the program objectives will provide a continuous pool of talented new contributors necessary for maintaining open-source projects. This study presents an empirical analysis of pull requests created by interns during the GSoC program. We extracted and analyzed 17,232 pull requests from 2,456 interns across 1,937 open-source projects. The results show most tasks involve both code-intensive activities like adding new features and fixing bugs, as well as non-code tasks like updating documentation and restructuring the codebase. Feedback from reviewers covers code functionality and programming logic, testing coverage, error handling, code readability, and adherence to best practices. Finally, we discuss the implications of these results for software engineering education.
\end{abstract}


\section{Introduction}
Most software engineering education research targets students in formal academic settings. Unfortunately, traditional approaches to software engineering education in these academic settings often limit students' exposure to and engagement with real-world projects, failing to fully harness their potential and creativity \cite{garousi2019aligning},\cite{garousi2019closing}. The industry frequently notes that fresh graduates have limited exposure to real-world projects and are often ill-equipped handle industrial challenges \cite{brechner2003},\cite{brng2013essence},\cite{ radermacher2013},\cite{radermacher2014}. Initiatives such as internships and university-industry collaborations on capstone projects have been designed to provide students with valuable industrial experience. However, these initiatives often require active industrial partnerships, which may not be always feasible.

Open-source software communities continuously need  new contributors to maintain and sustain their codebase. Contributions to open source can be publicly verified and may demonstrate competence during job searches. Open-source code repositories are abundant, easily accessible, and provide real-world examples that are essential to prepare students for the industry. Therefore, recruiting students to work on open-source code repositories equips the students with industrial experience, publicly-verifiable evidence of competence, and helps sustain the pool of contributors needed for open-source projects.

The Google Summer of Code (GSoC)\footnote{https://summerofcode.withgoogle.com/}\cite{hawthorn2008google} is an annual, global, three-month internship program that connects students to open-source software communities. Participants work on real-world open-source projects under the guidance of experienced mentors. Google provides stipends to student participants after completing certain milestones. The program is the first global initiative to introduce students to open source software development and bring in new contributors. Since 2005, GSoC has connected more than 19,000 new contributors from 112 countries with 18,000 mentors from 133 countries. The program has generated over 43 million lines of code across numerous projects for more than 800 open-source organizations\footnote{https://summerofcode.withgoogle.com/about}. The code contributions of the participants are publicly available, thereby providing valuable data to study novice activities in open-source development. 

%GitHub pull requests (PRs) serve as a critical mechanism for code contribution in software projects. 

Contributions to open-source are often made via pull requests where contributors clone a codebase, make changes (e.g, add a new feature) locally, and then notify the repository team that they have completed a feature or modified the codebase \cite{gousios2014exploratory}. Collaborators then review and provide comments on the proposed changes. Code review comments provide valuable feedback that guide the contributors and enhance code quality. Maintainers must approve the requested changes before they can be integrated into the main codebase. Although several studies have examined pull requests, there is limited research on the analysis of pull requests created by interns or students such as GSoC participants.

This study analyzes pull requests created by interns during the GSoC program from 2020 to 2023 (a 4-year period). We examine pull request data, review comments, project-related features, and contributors (intern) data to gain informative insights into the contributions interns make, the feedback they receive, and possible implications for software engineering education. The study analyzed 17,232 pull requests by 2,456 interns across 1,937 projects. We used  topic modeling techniques to gain insights into common themes in the pull request titles, descriptions and review comments. 
%We also used performed feature engineering to understand the features that contributes most to further contribution to the project outside of the GSOC program. Finally, we applied sentiment analysis on the post to extract the aggregate emotions in the posts.
In summary, the paper provides the following contributions:
\begin{enumerate}
    \item A publicly available dataset\footnote{https://github.com/compedutech/gsoc} of 17,232 pull requests created by GSOC interns from 2000 to 2023. This dataset was extracted directly from the official GSoC program page\footnote{https://summerofcode.withgoogle.com} (via web scrapping) and Github (via the GiHub API). The dataset also include the scripts used for web scrapping and extracting the pull requests from GitHub.
    \item Insights into the success or failure of the GSOC program in promoting open-source contributions. Specifically, we provide information on interns' continued contributions to the same project  after the program.
    \item An analysis of the tasks interns worked on and the feedback they received from the project collaborators.
    \item A discussion on the implications of these results for software engineering education. 
\end{enumerate}

%A number of research have highlighted the importance of internship in exposing students to real world project, enhancing their employability, and bridging the skill gap between academic courses and industry requirements. However, existing computing education pedagogy research have mostly focused on students within the academia.
\section{Background on Google Summer of Code Program}
Google Summer of Code (GSoC) is an annual, global program run by Google that encourages university students or new contributors to participate in open-source software development. The program pairs new contributors with mentors from open-source organizations. Google provide stipends to contributors who successfully meet some specified milestones. The following paragraphs provide a detailed overview of the program.
\begin{itemize}
    \item \textbf{Goal.} GSoC aims to introduce students to open-source software development by engaging them in real-world projects under the guidance of experienced mentors. The program has two main goals. Firstly, to expose students or novices to open-source development and help them gain real-world experience. Secondly, to provide open-source organizations with new contributors that will continue to contribute to open-source communities even after the program is completed.
    \item \textbf{Eligibility.} Participants must be students or new to open-source, at least 18 years old, eligible to work in their country of residence, and not residing in countries currently embargoed by the United States. The program is open to participants from all over the world who meet the criteria stated above.
    \item \textbf{Organization.} Open-source organizations apply to mentor students, and offer projects for students to work on. Organizations must have active open-source projects and not be based in embargoed countries.
     \item \textbf{Mentorship.} Experienced members of open-source organizations mentor students throughout the program.
     \item \textbf{Timeframe.} The standard timeline for the program is 12 weeks and it usually runs from May to August. However, mentors and contributors may agree to extend the program to 22 weeks.
     \item \textbf{Application process.} First, open-source organizations submit proposals for specific software development projects. Students would then create proposals explaining how they would approach the project. The mentoring organizations would then select students based on their proposals and demonstrated skills.
     \item \textbf{Onboarding and mentorship.}
     Selected students are paired with mentors from the open-source organization. The student spend the first few weeks to get familiar with the codebase and collaborate with mentors on expected project milestones. The students then spend the next twelve (or more) weeks working on implementing solutions and completing milestones. 
     \item \textbf{Completion.} Students deliver code, features, or bug fixes, and submit reports to both their mentors and Google.
     \item \textbf{Stipend.} Google provides stipends to students based on their location and duration of the program. The payment of the stipend are often contingent upon completion of some specified milestones.
     %\item \textbf{}
\end{itemize}
The GSoC program provides an excellent opportunity for students to develop their skills and make meaningful contributions to projects used by people around the world. There are numerous benefits associated with the program and they include the following \cite{silva2020google},\cite{tan2023understanding}.
\begin{enumerate}
    \item \textbf{Learning opportunity.}  Students gain hands-on experience in software development in a real-world context.
    \item \textbf{Networking.} Participants build relationships with mentors and other open-source contributors. 
    \item \textbf{Career development.} The contributions implemented during the GSoC program are public and easy to verify. Many students saw the GSoC program as an opportunity to build their resume, kickstart their career, and increase their visibility when seeking a job \cite{silva2020google}. Furthermore, the program is popular, highly regarded by employers, and many students go on to work for the organizations they contributed to. This makes participation in the program an excellent component of participants' resume when searching for jobs. 
    \item \textbf{Open-source contribution.} Participants contribute code to open-source projects, which is publicly available and can be used by the global software community.
    \item \textbf{Open-source sustainability.} The program provides new contributors to open-source communities which is necessary for sustaining the open-source projects. Tan et al. \cite{tan2023understanding} noted that sustainability of the project was a major motivating factor for mentors to participate in the program. 
\end{enumerate}
The GSoC program offers an excellent avenue to introduce students or newcomers to open source, and support the sustainability of open-source communities. Majority of the organizations participating in the program have also reported that the program helps them to find new contributors who are active and committed to the open source project \cite{hornig2019summer}. The contributions of students to the open source community are also public and accessible to anyone. This public data from students' contributions provides a valuable dataset for analyzing the program's impact, understanding challenges or barriers to open-source software development, and probe factors that ensure successful engagement with the open source communities.
\section{Literature Review}
Several studies have examined the impact of summer coding programs, internships, and pull requests on software development and education. However, to the best of our knowledge, none of the studies have provided empirical analysis of pull requests created during summer programs. Silva et al. \cite{silva2020google}\cite{silva2020theory} conducted a qualitative study to understand students' motivations for participating in GSoC program, and they then developed a theoretical framework to model students' engagement. Their findings reveal that students often joined the program to acquire new software engineering skills, enhance their resumes, and earn stipends. Tan et al. \cite{tan2023understanding} explored mentors' motivations in the GSoC program and noted that intrinsic motivation to sustain the open-source codebase, learn new skills, and increased recognition in the open-source community are the major reasons why mentors participate in the program. Trainer et al. \cite{trainer2014community} analyzed 22 GSoC projects and 22 hackatons to understand student engagements with the open-source scientific community. Their findings reveal that private mentor-student communications often create strong ties, while public sharing of progress reports create weak ties with other community members. Interestingly, 18\% of students later became mentors. 

Majority of the approach so far have used qualitative study (without considering the codebase) to analyze students and mentors participation in the GSoC program. The closest to our work was conducted by Silva et al. \cite{silva2017students}\cite{silva2017long} who analyzed the code commits of 866 students in the GSoC program from 2013 to 2015. The authors discovered a strong correlation between the number of commits and students' retention within projects during and after the GSoC program. Additionally, 82\% of the projects merged at least one commit from students into their codebases. However, while commits provide granular insights into code changes, they may not fully capture task types or objectives. Pull requests on the other hand, contains aggregate changes from multiple commits and provide a high-level data on the kind of tasks the students work on. Pull requests also provide opportunities for other members of the community to review changes, provide comments or feedback, and suggest modifications to the proposed changes \cite{gousios2014exploratory}. Furthermore, pull requests have been shown to be more precise than commits in evaluating the contributions of community members \cite{bertoncello2020pull}. Our study extends the work by Silva et al. \cite{silva2017students}\cite{silva2017long} to use pull requests for understanding students' contributions to the GSoC program.

%unmerged but closed pull request from Gousios et al.
Studies on pull requests have revealed their significance in collaborative software development. Gousios et al. \cite{gousios2014exploratory} conducted an exploratory study of the pull-based model on 291 projects. They found out that the pull-based model reduces the time needed to merge contributions to the codebase, enhances development turnover, and encouraged community engagement. Kalliamvakou  et al. \cite{kalliamvakou2014promises} also conducted a study on GitHub repositories and noted that pull requests are often used for code reviews and issue resolutions. Liu et al \cite{liu2016comparative} reported that pull requests can lead to more active code repositories. Yu et al. \cite{yu2014reviewer} and Kerzazi et al. \cite{kerzazi2016can} developed theoretical frameworks for recommending pull requests' reviewers. Zhang et al. \cite{zhang2018multiple} examined competing pull requests and noted that 31\% of pull requests in 75\% of repositories tend to modify the same code during the same period. Jiang et al. \cite{jiang2019characteristics} examined reopened pull requests and noted that reopened pull requests often had more comments and lower rate of being merged with the codebase. Legay et al. \cite{legay2018impact} examined the impact of the acceptance or rejection of a pull request on future contributions to a project. Their results show that contributors with higher acceptance rate are more likely to continue contributing to the project, while those with rejections are less likely to contribute in the future. The studies discussed in this paragraph have examined pull requests in different contexts. However, none of the studies have examined pull requests created by novice programmers or students. This study extends the literature to provide an in-depth analysis of pull requests from students or new contributors in the GSoC program.

Internships and industry-based capstone projects play crucial roles in providing students with real-world exposure. Kapoor et al. \cite{kapoor2019} and Lehman et al. \cite{lehman2024sealing} found that internships significantly improves professional development, while Groeneveld et al. \cite{groeneveld2019} noted that both internships and capstone projects promotes the development of soft skills among students. Dean et al. \cite{dean2011} emphasized that the impact of internships on students varies by student involvement levels. Sudol et al. \cite{sudol2010analyzing} reported that internship experience reduces misconceptions about software engineering concepts. The related work on internships discussed so far have adopted qualitative studies to analyze the impact of internships on students without considering the actual code contributions. The closest work to our study was by Menezes et al. \cite{menezes2022open} where they described their experience on an internship model that pairs students with volunteer mentors in open-source projects. The authors then conducted a qualitative study using surveys and reports to understand the impact of the internship model on students' skills and job placement. Their results show significant increase in students' technical skills, soft skills, and job placement after graduation. They also noted that students who are paid stipends are more willing to work full time on the project. 

This literature review section have discussed studies related to GSoC, pull requests and internships. Although, these studies provided valuable insights, none of them conducted large-scale analyses of pull requests within the context of internships or summer coding programs. This study fills this gap by examining GSoC participants' pull requests, to offer a comprehensive view of their contributions and feedback.

\section{Design Methodology}
This section outlines the data collection process, research methodology, and the research questions we aim to answer.
\subsection{Data Collection}
We extracted pull request data using web scrapping and the GitHub API. The following paragraphs highlights how we extracted the needed data.
\begin{enumerate}
    \item The Selenium Python Package was used to scrape the URLs of each intern's personal reflection website on the GSoC web page for the years 2020 to 2023.
    \item The personal reflection websites of the GSoC participants often contained links to their contributions during the program. So, we extracted all URLs from each intern's webpage by locating HTML anchor tags.
    \item URLs containing "GitHub" and either "issues" or "pull" were filtered to identify pull request links. For example the url link "https://github.com/repos/asyncapi/community/pulls/720" contains the words "github" and "pull". Most pull requests URLs on GitHub often follow this pattern and we believe it is the easiest way to extract most of the pull request links.
    \item The GitHub API was used to extract the pull request data associated with the filtered URLs.
\end{enumerate}
The dataset includes three main categories: pull request details, contributor (intern) information, and the project metadata. The final dataset comprises of  17,232 pull requests submitted by 2,456 interns across 1,937 open-source projects. The following subsections provides statistical overview of the pull requests, projects and users (contributor) information embedded in the dataset.
\subsubsection{Pull Requests Data}
The pull request dataset contains 22 features that captured different aspects of the pull requests. Key features include the title (title of the pull request), body (description of the pull request), is\_merged (if the pull request has been merged to the code base), created\_at, closed\_at, review\_comments, commits, and so on. We also computed a derived feature from the the original data for additional insights. The derived feature is called "time\_taken" and it captures the duration between opening and closure of a pull request.

The final pull request dataset encompasses 17,232 pull requests involving 60 million lines of code added, 18 million lines deleted, and modifications to over 300,000 files. On average, a pull request remained opened for 4.7 hours, garnered 4.3 comments and received 9.5 review comments. Table \ref{table_pull_stat} provides a statistical description of key pull request features. The is\_merge feature is assigned a value of 1 if the pull request has been merged and a value of 0 otherwise. Similarly, the state feature is assigned a value of 1 if it is closed, else it is assigned a value of 0. The table shows that 85.7\% (14,775) of the pull requests were successfully merged with the codebase while 94.7\% (16,235) were resolved and closed.

\begin{table*}[t]\centering
    \begin{tabular}{|r |r |r|r |r|r|r |r|r|r |r|}
        \hline
        \multicolumn{1}{|c|}{Statistics}  & 
        \multicolumn{1}{c|}{Is\_merged} & \multicolumn{1}{c|}{State} &
        \multicolumn{1}{c|}{Additions} & \multicolumn{1}{c|}{Deletions} &
        \multicolumn{1}{c|}{Changed Files} & \multicolumn{1}{c|}{Commits} &
        \multicolumn{1}{|c|}{PR comments}  & \multicolumn{1}{c|}{Review Comments} & \multicolumn{1}{c|}{Time taken (Hours)}  \\
        \hline
        Total & 14775 & 16325 & 60,049,227 &	18,668,966 &	353,946 &	198,880	&74, 343 &	165,330 & 14,898,582 \\
         \hline
        Mean & 0.857 & 0.947 & 3484.95 &	1083.45 & 20.5 & 11.5	&4.3 &	9.5 & 4.7 \\
         \hline
        Std Dev & 0.35 &0.22& 67352.30	&52568.90&	263.91 &84.04 & 8.48 &26.08 & 2812\\
         \hline
    \end{tabular}
    \vspace{0.3cm}
    \caption{Descriptive Statistics of Pull Request}
    \label{table_pull_stat}
\end{table*}

\begin{table*}[t]\centering
    \begin{tabular}{|r |r |r|r |r|r|r |r|r|r |}
        \hline
        \multicolumn{1}{|c|}{Statistics}  & 
        \multicolumn{1}{c|}{Forks} & \multicolumn{1}{c|}{Stars} &
        \multicolumn{1}{c|}{Watchers} & \multicolumn{1}{c|}{Contributors} &
        \multicolumn{1}{c|}{Commits} & \multicolumn{1}{c|}{Issues} &
        \multicolumn{1}{|c|}{Branches}  & \multicolumn{1}{c|}{Pull Requests}  \\
        \hline
        Total & 1104727 & 3479159 & 117861 &	335513 & 14141978 &	328060	&66728 &	53210 \\
         \hline
        Mean & 570.62 & 1797.09 &	60.88 &	173.30&	7304.74&	169.45& 34.47 &	27.48 \\
         \hline
        Std Dev & 3132.37 &	8642.19 &	265.01 &	794.45 &	34255.74 & 864.12 & 142.84 &	131.92\\
         \hline
    \end{tabular}
    \vspace{0.3cm}
    \caption{Descriptive Statistics of Projects }
    \label{table_project_stat}
\end{table*}

\begin{table*}[t]\centering
    \begin{tabular}{|r |r |r|r |r|r|r|r|r|}
        \hline
        \multicolumn{1}{|c|}{Variables}  & 
        \multicolumn{1}{c|}{Number of Pull} & \multicolumn{1}{c|}{Accepted Pull} &
        \multicolumn{1}{c|}{Reject Pull} & \multicolumn{1}{c|}{Projects} &
        \multicolumn{1}{c|}{Public Repositories} &
        \multicolumn{1}{|c|}{Following}  & \multicolumn{1}{c|}{Followers}  \\
        \hline
        Total & 17232 & 14775 & 2457 &	3117 & 91957 &	66295	&124813  \\
         \hline
        Mean & 8.28 &	7.10 &	1.18 & 1.50 &	45.32 &	32.67 &	61.51 \\
         \hline
        Std Dev & 10.37 & 9.72 & 2.83 &	1.54 & 47.30 &	76.84 &	287.95\\
         \hline
    \end{tabular}
    \vspace{0.3cm}
    \caption{Descriptive Statistics of Users }
    \label{table_user_stat}
\end{table*}

\subsubsection{Projects Data}
\begin{figure}
    \centering
    \includegraphics[width=0.9\linewidth]{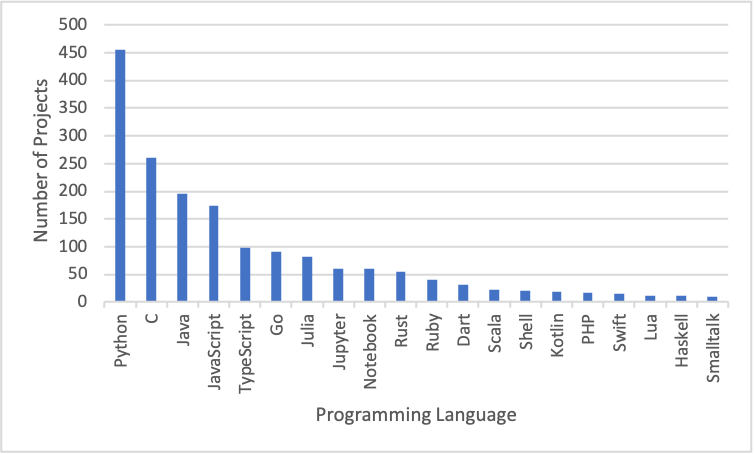}
    \caption{Number of Projects associated with each Programming Language}
    \label{fig_project_langauge}
\end{figure}

The project dataset includes metadata for 1,937 projects, with features such as project name and programming language. Other features are number of Forks, Stars, watchers, contributors, commits, branches, issues, and pull requests. Table \ref{table_project_stat} summarizes these details, and highlights the popularity and complexity of these projects, which averaged 1,795 stars and 34 branches. The top ten projects based on the number of stars \cite{ray2014large} are
tensorflow/tensorflow, huggingface/transformers, nodejs/node,
rust-lang/rust, godotengine/godot,
django/django, opencv/opencv ,tensorflow/models, swiftlang/swift, and
webpack/webpack. Python was the dominant programming language, as shown in Figure \ref{fig_project_langauge}, which visualizes the number of projects associated with the top 20 languages.

\subsubsection{Users Data}
The user dataset captures information related to each intern that participated in the GSoC program. The user dataset captures 9 features including the contributor id, the number of projects they worked on during the GSoC program, the number of pull requests created during the GSoC program, number of accepted pull requests,  number of rejected pull request, and all the review comments associated with the user. Other features include the number of public repositories, followers and followings.

There were a total of 2456 interns in the dataset, including 154 duplicates. The duplicates were interns who participated in multiple sessions of the GSoC program (e.g., a student who participated in the 2020 and 2021 programs). For this study, the duplicated users were treated as new users because the user contributions (e.g number of pull requests created) will differ from year to year. Table \ref{table_user_stat} summarizes users statistics, noting that the aggregate number of projects reflects cumulative contributions rather than unique project counts (this was captured in Table \ref{table_project_stat}). These metrics provide a comprehensive view of user engagement and impact during GSoC.

\subsection{Research Questions}\label{rq}
The goal of this study is to provide insights into the pull request activities created by interns during the GSoC program. A detailed analysis of the pull requests can reveal patterns in the contributions interns make to open-source repositories. It may also shed light on how collective thinking emerges among a group of individuals, the core features interns often work on, the type of feedback they frequently receive, and factors that contribute to a successful or unsuccessful GSoC experience. These insights can inform necessary changes to the software engineering curriculum, encourage students to contribute to open source, and better prepare students for the workforce. In this context, this study is guided by the following research questions.

%What are the major characteristics of the project , users (interns), and pull request - Descriptive statistics -154 users had multiple programs
\begin{itemize}
    \item \textbf{RQ1. How successful has the GSoC program been in encouraging contribution to open source communities?} A continuous stream of new contributors is crucial to the sustainability of the open-source projects. One of the major goals of the GSoC program is to encourage contributions to open-source. This research question aims to evaluate how effectively the GSoC program attracts and retains new contributors.
    \item \textbf{RQ2. What are the major tasks that interns often work on?} This question seeks to categorize the tasks or contributions interns commonly engage in. This can provide insights into interns' interest and lay a foundation for attracting new comers to open-source communities.
    \item \textbf{RQ3. What are the major feedback interns often receive on their contributions.} Pull request contributions are typically reviewed by experienced contributors or moderators, who provide comments on how to improve the contributions. These feedback can offer valuable insights into the challenges interns face, identify patterns of mistakes or errors, and inform improvements to the software engineering curriculum to better prepare students for the workforce and open-source contributions.
\end{itemize}
These three research questions aims to deepen our understanding of interns' contributions to open-source projects, provide guidelines for retaining new contributors, and discuss the implications of these findings for software engineering education.

\subsection{Study Design}\label{design}
For  the study, we used topic modeling techniques to address the research questions. Topic modeling, or topic clustering, is an unsupervised machine learning method for discovering the underlying thematic structures in textual content. This process generally consists of three key steps: pre-processing, topic modeling, and interpretation of results.

Firstly, we pre-process the text documents to improve the accuracy of topic modeling. Texts were converted to lowercase and cleaned to remove irrelevant characters (such as '/'), punctuations, numbers, and short words with only one or two letters. Then, we applied tokenization to break the text into individual words, followed by removal of stop words (e.g., 'a', 'an' 'the'), and stemming to reduce words to their root forms (e.g., converting 'understanding' to 'understand').

Secondly, the Latent Dirichlet Allocation (LDA) \cite{blei2003latent}, implemented using the Gensim Python library \cite{rehurek2011gensim}, was employed to identify topics in the document. Coherence scores were used to compare a range of topic numbers in order to select the optimal number of topics for each tasks.

Finally, we interpreted the results by associating the generated topics with their respective keywords. We selected posts that contains the identified keywords, manually reviewed the posts to understand the context, and then applied open-coding techniques to label the topics generated from the Gensim package. We adopted a 2-step open-coding process where we first extract the top 20 keywords for each topic identified by the LDA algorithm and then manually assigned labels to the topics based on these keywords. The goal of this final interpretation process is to ensure that we generate meaningful descriptions for the main identified topics.

To answer the first research question, we extracted all the pull requests created by a user in a project. We then separated the pull requests that were created by that user before or during the GSoC program. This allows us to examine the contributions of the user to the project after the GSoC program. It should be noted that we were able to access only the pull requests created by the user in the project(s) they worked on during the GSoC program and not other projects. We were unable to track if the user contributed to other open-source projects outside of the project they worked on during the GSoC program.

To answer the second research question, we applied topic modeling techniques to the title and body of the pull request. The titles convey the purpose and scope of the changes while the body contains the descriptions of the pull requests. The body of a pull request is crucial for providing context and explaining the changes made in the code, allowing reviewers to understand the scope and rationale behind the work. A well-crafted pull request body typically includes a clear explanation of what the pull request does, the reason for the change, and any special instructions for testing or deployment. Therefore, a detailed analysis of pull request titles and bodies (or descriptions) allows us to understand the major tasks or topics the interns worked on.

To answer the third research question, we applied topic modeling to the review comments. The review comments in the dataset contains valuable feedback provided to the interns. These comments are essential for improving the code quality, ensuring adherence to best practices, and enhancing clarity and maintainability. A well-constructed review comment provides constructive criticism, suggests improvements, and sometimes includes code snippets to guide developers. An analysis of these comments can identify key areas where interns frequently encounter challenges and need guidance.
This methodology ensures a structured analysis of interns’ contributions, tasks, and feedback, thereby offering valuable insights for improving software engineering education and supporting open-source community engagement.
%sentiment analysis of comments
%topics of title and description -> What topics or area do they interns work on

%features correlated to further contribution

%topics of review comments - What are the major pull request contribution or feedback or 

\section{Results}
This section presents the results of our analysis with respect to the three research questions discussed in Section \ref{rq}.
\subsection{RQ1: GSoC Success}
The first research question examines how successful the GSoC program is in promoting open-source contributions. The results of the analysis from Table \ref{table_pull_stat} show that 85.74\% (or 14,775 out of 17,232) of the pull requests were accepted and successfully merged to the codebase. Furthermore, 1,113 out of 2,456 users had all their pull requests accepted while 319 had no accepted pull request. 

Next, we examined the pull requests created by each user before and after the GSoC program. The results show that 63.5\% (1,561 users) contributed at least one pull request to the same project after the program, while 70.7\% (1,738 users) contributed at least one pull request to the project before starting the GSoC program. This shows that the majority of users started contributing to the project before the GSoC program and many continued to do so after the program. In fact, only 15\% contributed solely during the GSoC program with no contributions before or after the program. This may indicate that the GSoC program motivates students to contribute to open source. Students may also have contributed to an open source project before the program to increase their chances of being selected.
%718 never contributed to the project before the program while 895 never contributed after the program. 375 contributed only during the program (neither before or after)

\subsection{RQ2: Tasks and Contributions}
\begin{figure}
    \centering
    \includegraphics[width=0.9\linewidth]{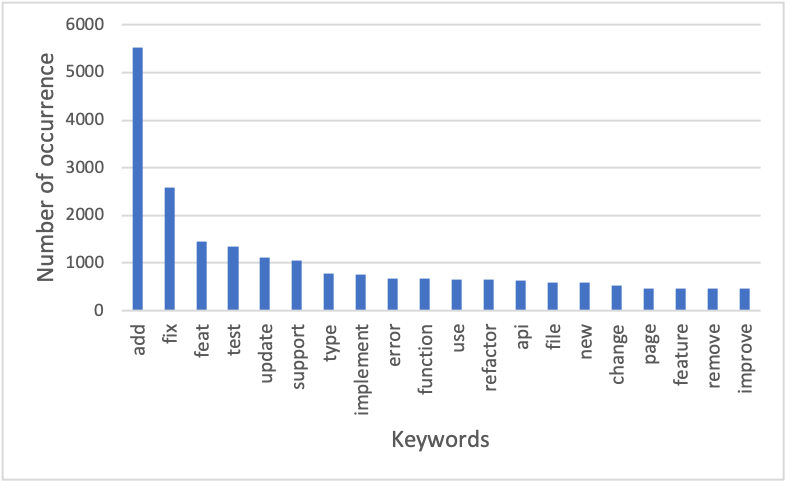}
    \caption{Frequencies of Words in the Title}
    \label{fig_word_title}
\end{figure}

\begin{figure}
    \centering
    \includegraphics[width=0.9\linewidth]{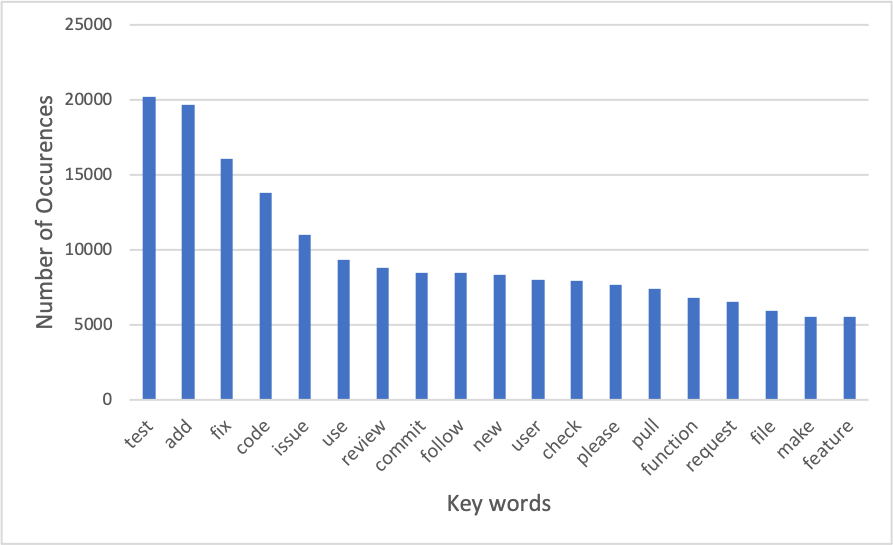}
    \caption{Frequencies of Words in the Body (Description)}
    \label{fig_word_body}
\end{figure}

We used word frequencies and topic modeling techniques to analyze the titles and descriptions (body) of the pull request with the goal of extracting the types of tasks students worked on. First, we extracted the most frequent words that appear in both the titles and descriptions. Figure \ref{fig_word_title} and Figure \ref{fig_word_body} illustrates the most recurring word in the titles and descriptions.


Next, we used LDA topic modeling to extract the major topics in the descriptions and the key words that capture these topics. Thirdly, we extracted sample text from the documents containing the keywords from the word frequency and topic modeling in order to understand the context of the words. Finally, we categorized the topics based on the keywords and the context in which they appear. The extracted topics for the tasks are.
\begin{enumerate}
    \item \textbf{Adding new features and APIs.} This topic involves developing and implementing new features like knowledge panels or API endpoints. Sample pull requests with this topic include: ``Added support for opening a file", ``Added custom IDE and improved other IDE launch commands", and ``Add an explore mode to run the app without marking messages as read".
    \item \textbf{Corrective maintenance tasks.} This topic covers pull requests that attempt to correct an error or ensure a feature is working as expected. Examples of pull requests in this topic include ``Fix double-checked locking in ConnectionFactory", ``Fix axeDev tools + lighthouse error for learner dashboard and make the lighthouse score to 1", and ``Add and update tests".
    \item \textbf{Project setup.} This topic covers pull requests that deal with initializing the project structure and adding necessary dependencies. Examples of pull requests in this topic include: ``Setting up the foundational infrastructure for FastAPI", ``Code relocated under maryam/", ``added setup.py, src: setup: Add root install support", ``Initial PlanetGenerator Setup", and ``Created the spring boot project with the necessary dependencies".
    \item \textbf{Documentation updates} This topic covers pull requests that relate to making changes to the README file or other documentation files to ensure clarity and up-to-date information. Sample pull requests include "Updating or adding information to ReadMe files",  ``Privacy Statement Updates September 2022", ``Added basic chondro documentation", ``restructured install documentation".
    \item \textbf{Refactoring.} This topic covers pull requests that aim to change parts of the codebase without changing behavior. Sample topics include ``wxGUI refactoring: New WorkspaceManager class", ``refactor: color blending function", ``refactoring of webapp to consume typescript client", and ``[GSoC] API change for the newly added system of ODEs solvers".
\end{enumerate}

The analysis of the second research question shows that while majority of the contributions are code-intensive, a significant portion of the pull requests involves changes to the documentation files and restructuring the organization of the codebase. This may also indicate the need to emphasize the variety of contributions student can make to open source.

\subsection{RQ3: Feedback and Review Comments}
To answer the third research question, we also applied word frequencies and topic modeling techniques to the review comments to understand the type of feedback reviewers provide for the pull requests. The topics identified in the review comments include the following.
\begin{enumerate}
    \item \textbf{Code organization and structure improvements.} This topic emphasizes project structure and organization, particularly where code files, tests, and other assets should be placed. This helps to maintain a clean project hierarchy, improve readability, and simplify navigation for future developers. Sample comments include ``You should put test in a tests/ directory" and ``If you move this to the test directory, it becomes clearer".
    \item \textbf{Best practices and coding style.} This topic includes comments that suggest changes that align with best practices, such as ensuring minimal exposure of unnecessary components and adherence to consistent coding styles. This helps ensure code reliability, consistency, and security. Sample comments include ``I always prefer if we keep safe and only publish necessary things" and ``Why not just use a str for facet\_value?".
    \item \textbf{Functionality and logic corrections.} This topic relates to the clarity of logic or ensuring the code behaves as intended. Reviewers help identify sections where functionality may not align with the expected outcome or where the logic could be improved. Sample comments include ``This is more the why of knowledge panels" and ``The logic here could be improved by changing this function" 
    \item \textbf{Testing coverage and validation.} This relates to adding more tests or moving test cases to appropriate locations to ensure the functionality is well-covered. They emphasize the importance of testing edge cases and ensuring that new features are adequately validated before being merged. Sample comments include ``You should write a test for this specific scenario" and ``This becomes easier to test if you move it into a dedicated test case"
    \item \textbf{Error handling and resilience.} This topic focuses on ensuring that the code is resilient to errors and can handle potential failure cases gracefully. Reviewers point out where additional checks or error handling mechanisms can be implemented to improve robustness. Sample comments are ``Make sure we handle this case properly when the data is missing" and ``Consider adding a check for this specific error".
    \item \textbf{Code readability and clarity.} This topic relates to requests for improvements in code readability, suggesting changes that make the code more understandable. This involves better naming conventions, simpler logic flows, and clearer documentation or comments. Sample review comments include ``This part of the code could be written more clearly" and ``The variable names should better reflect their purpose".

\end{enumerate}

%Suggested Improvement: Follow standard conventions for organizing directories and files (e.g., placing tests in a dedicated tests/ folder) to ensure the project remains maintainable and scalable.

%Suggested Improvement: Maintain consistency in data types, error handling, and security considerations. Always ensure that only the necessary components are exposed or published and use standard practices to avoid technical debt.

%Suggested Improvement: Clarify the intent of functions and ensure the logic follows best practices. Refactor code to make logic more straightforward and robust to minimize potential bugs.

%Suggested Improvement: Increase testing coverage by adding tests for various scenarios, including edge cases and expected failures. Refactor tests to improve clarity and modularity, ensuring that tests align with the overall project structure.

%Suggested Improvement: Implement proper error handling and validation checks throughout the code to ensure it can deal with unexpected inputs or conditions. Use specific try-catch blocks or validation methods to handle these scenarios gracefully without crashing.

%Suggested Improvement: Focus on writing clean, readable code with meaningful variable names and concise comments where needed. Make sure that other developers can easily understand the code without additional explanations.

The result presented above shows some key areas that reviewers have focused on. It can be observed that reviewers are not only concerned about the correctness of the code, but are also interested in ensuring that the newly added or modified code is easy to maintain.

\section{Discussion and Implications for Software Engineering Education}
The results of the analysis on GSoC pull requests and review comments have significant implications for software engineering education. The identified tasks, and feedback from review comments can help educators better prepare students for real-world software development practices. The findings highlight key areas that should be emphasized in software engineering curricula, as outlined below.
\begin{enumerate}
    \item \textbf{Importance of code reviews in collaborative software development.} The study underscores the critical role of peer reviews in ensuring the quality of code contributions. For software engineering education, this suggests integrating code reviews into course projects and emphasizing communication skills to facilitate effective collaboration. 
    \item \textbf{Code organization and structure.} The findings show that reviewers frequently comment on the structure and organization of code, such as file placement and module organization. This highlights the need to emphasize code organization in software engineering education by 1) teaching modular design and file organization, and 2) following real-world practices in project layout within software engineering courses. 
    \item \textbf{Testing and test-driven development} The study reveals that a significant portion of review feedback pertains to missing or inadequate testing. Educators should therefore place greater emphasis on teaching comprehensive testing practices, including test-driven development and  addressing edge cases.
    \item \textbf{Error handling and robustness.} Review comments often address issues with error handling, suggesting that students may underestimate its importance. Educators should teach best practices for error handling and include this aspect as part of assessment criteria.
    \item \textbf{Performance optimization.} Reviewers also suggest performance optimizations or improvements in coding logic. For software engineering education, this highlights the importance of teaching algorithmic efficiency and introducing profiling tools to identify performance bottlenecks.
    \item \textbf{Refactoring and continuous improvement.} Reviewers often recommend refactoring code for improved readability, efficiency, and maintainability. This underscores the need to encourage refactoring as a regular habit in software engineering education while teaching clean code principles.
    \item \textbf{Preparing Students for open-source contributions.} Many findings reflect practices commonly observed in open-source development environments, such as GitHub. The alignment of software engineering education with these practices can better prepare students for contributing to open-source projects. Educators can use real-world GitHub projects as teaching tools, and incorporate version control systems to stimulate collaboration in software engineering classes.
    \item \textbf{Learning to give and receive feedback.} The study highlights the value of specific and constructive feedback in review comments. Software engineering education should emphasize 1) the ability of students to provide constructive feedback and 2) the importance of receiving and acting on feedback from peers.
\end{enumerate}

The findings of this study offer several key insights for software engineering education. When educators focus on real-world practices such as code organization, testing, error handling, performance optimization, and collaboration; they can better prepare students for the challenges of the software industry. The integration of code reviews, refactoring exercises, and open-source contributions into the curriculum will likely foster a culture of continuous improvement and collaboration. This approach equips future software engineers with the skills and mindset they need for success.  These educational adjustments will help students internalize best practices in software development, become better collaborators, and emerge as thoughtful, effective developers in both open-source and proprietary software environments.
\section{Threats to Validity}
There are several potential threats to the validity of our findings. These threats can affect the reliability and generalization of the results. The following subsections discuss the potential threats to the validity of the results in terms of construct validity, internal validity, external validity, and conclusion validity.

\subsection{Construct Validity}
Construct validity refers to whether the measures used in the study accurately capture the concepts or phenomena being investigated. In this case, the key constructs are the themes extracted from pull request titles, descriptions, and review comments. The following represent the construct validity threats.
\begin{itemize}
    \item \textbf{NLP limitations in identifying themes.} The natural language processing (NLP) techniques used in the study, such as keyword extraction, tokenization, and topic modeling may not fully capture the nuances and context of review comments. For instance, comments that use implicit language, technical jargon, or sarcasm might be misinterpreted by keyword-based approaches. To minimize this threat, we manually validate a sample of comments. However, incorporating more advanced models (e.g., transformer-based models) could improve the accuracy of capturing nuanced feedback.
    
    \item \textbf{Pull request descriptions may not fully capture the nature of changes.} Pull request descriptions vary widely in quality and completeness. Some pull request may lack detailed explanations of the changes, leading to incomplete or inaccurate summaries of tasks.
%Mitigation: Encouraging more structured PR descriptions through guidelines or automated checks could help improve the quality of PR metadata.
\end{itemize}

%Misclassification of review themes: The process of categorizing feedback into themes (e.g., code structure, testing, performance) may oversimplify complex comments. For example, a comment mentioning both functionality and performance issues may be classified into only one category, missing the multifaceted nature of the feedback.
%Mitigation: A multi-label classification approach could be implemented where a comment can belong to multiple categories if necessary. Additionally, human validation of the categorized comments can reduce misclassification errors.

\subsection{ Internal Validity}
Internal validity refers to the extent to which the observed results can be attributed to the research methods used, rather than external factors. This includes issues that may arise from the dataset or the methods of analysis.
\begin{itemize}
    \item \textbf{Bias in data collection.} The dataset used for the study was sourced solely from GitHub repositories of projects that participated in the GSoC program.  Since this program in being run by only Google, there is a possibility that the projects follow specific coding practices, use particular technologies, or operate in niche domains. Hence, the findings may not be representative of broader software development practices.
    \item \textbf{Incomplete review comments.} Reviewers may not comment on all issues in the code due to time constraints or priorities. As a result, important problems or suggestions for improvements may be overlooked in the analysis.
\end{itemize}

%Mitigation: To reduce this bias, a larger, more diverse dataset of PRs from a variety of projects, domains, and communities could be collected and analyzed. Alternatively, focusing on popular repositories with diverse contributors could provide more generalizable insights.
%Selection Bias in Review Comments: Review comments analyzed in the study are often written by more experienced developers or maintainers. These individuals might offer feedback that is more detailed or rigorous compared to less experienced contributors. Consequently, the analysis may overestimate the quality or depth of typical code review comments.
%Mitigation: Stratifying the analysis by the experience level of the reviewers could offer insights into how feedback changes with experience and help ensure that a balanced view of review quality is represented.
%Incomplete Review Comments: Reviewers may not comment on all issues in the code due to time constraints or priorities. As a result, important problems or suggestions for improvements may be overlooked in the analysis.
%Mitigation: Complementing the analysis of review comments with automated static analysis tools (e.g., linters, code quality checkers) can help identify issues that reviewers might have missed. Comparing human feedback with automated feedback could provide a fuller picture of code quality concerns.
\subsection{External Validity}
External validity pertains to the generalizability of the findings. This involves assessing whether the results of the study can be applied to other software projects, development teams, or domains.

\begin{itemize}
    \item \textbf{Generalizability across different projects and domains.} The study may not generalize well to all software projects. Different types of software projects (e.g., open-source vs. proprietary, small vs. large projects, front-end vs. back-end development) have distinct workflows, contributor behaviors, and review cultures. Therefore, findings from a single type of project may not apply broadly.
    \item \textbf{Varying review practices across teams.} Review practices may differ based on team norms, the development process (e.g., agile vs. waterfall), or company policies. A highly structured review process may yield different types of feedback compared to a more informal review process.
    \item \textbf{Changes over time.} Feedback practices and development workflows on GitHub may evolve over time, influenced by the introduction of new tools (e.g., code linters, continuous integration pipelines) or changes in the development community's culture. This temporal aspect might limit the applicability of findings from historical pull request data to current or future practices.
\end{itemize}
%Mitigation: A wider selection of repositories from different domains, programming languages, and project sizes would improve generalizability. Additionally, comparing open-source and proprietary projects might reveal differences in review practices and feedback.
%Varying Review Practices Across Teams: Review practices may differ based on team norms, the development process (e.g., agile vs. waterfall), or company policies. A highly structured review process may yield different types of feedback than a more informal review process.
%Mitigation: Examining PRs from teams with different review processes, or analyzing a range of PRs from diverse teams, would help address this threat. Explicitly studying how the review process itself impacts the nature and quality of feedback could further improve generalization.
%Changes Over Time: The feedback practices and development workflows on GitHub may evolve over time, influenced by the introduction of new tools (e.g., code linters, continuous integration pipelines) or changes in the development community's culture. This temporal aspect might limit the applicability of findings from historical PR data to current or future practices.
%Mitigation: A longitudinal study that examines how PR feedback evolves over time could shed light on temporal shifts in development practices. Additionally, comparing findings from recent PRs with older ones could help identify trends or changes in code review patterns.
\subsection{Conclusion Validity}
Conclusion validity concerns whether the conclusions drawn from the analysis are justified by the data and methodology used.

\begin{itemize}
    \item \textbf{Over-interpretation of statistical results.} While keyword frequency and thematic analysis offer valuable insights, there is a risk of over-interpreting correlations as causal relationships. For example, a high frequency of comments mentioning "testing" might not necessarily imply that testing is a major issue across all pull request, but could reflect a focus on testing in certain projects.
    \item \textbf{Unaccounted confounding factors.} Confounding factors such as the complexity of the code changes, the experience level of the contributor, or the urgency of the pull request could influence the nature of the feedback provided. If these factors are not accounted for, the conclusions may be skewed.
\end{itemize}

%Mitigation: Cross-referencing the results with expert judgment or triangulating findings with additional qualitative analysis can help validate the interpretations. Involving subject matter experts in the evaluation of the themes could also ensure that the conclusions drawn are meaningful and contextually appropriate.

%Mitigation: Controlling for these confounding factors, such as by stratifying PRs based on their size or the contributor's experience, would improve the validity of the conclusions. Moreover, conducting multivariate analysis could account for the influence of multiple factors at once.

The study provides valuable insights into pull requests and code review comments. However, the validity of the findings could be affected by the aforementioned threats. We have used various approaches (such as manual validation of some of the comments) to minimize these threats and ensure careful interpretation of the results. In the future, we will expand the dataset, refine the analytical techniques, and account for variability in development practices to enhance the validity and generalizability of the findings.
\section{Conclusion}
This study provides an in-depth analysis of GitHub pull requests and review comments created during the Google Summer of Code program from 2020 to 2023. The study offers insights into common tasks and feedback themes. The results shows that pull requests frequently focus on feature additions, bug fixes, and documentation updates. The review comments often emphasize the need for improved code organization, additional tests, performance optimizations, and better error handling.

These findings can help developers and teams enhance their review practices by focusing on key areas such as test coverage, performance improvements, and maintaining a well-structured codebase. The results also highlights the need to enhance the software engineering curriculum by incorporating a stronger emphasis on code reviews, code organization, testing, and best practices from open-source repositories.

In the future, we plan to expand the dataset to include data for all available years of the GSoC program. Additionally, we aim to study the evolution of tasks and comments across these years. Finally, we intend to explore other summer programs and hackathons that target students to validate the generalizability of this study's findings across different repositories and software domains.

\newpage
\bibliographystyle{elsarticle-num} 
\bibliography{ref}

\end{document}